\documentstyle[12pt,a4]{article}
\newcommand {\beq}{\begin{equation}}
\newcommand {\eeq}{\end{equation}}

\newcommand{\be}{\begin{equation}}
\newcommand{\ee}{\end{equation}}
\newcommand{\bea}{\begin{eqnarray}}
\newcommand{\eea}{\end{eqnarray}}
\newcommand{\bean}{\begin{eqnarray*}}
\newcommand{\eean}{\end{eqnarray*}}

\newcommand{\gapproxeq}{\lower .7ex\hbox{$\;\stackrel{\textstyle >}{\sim}\;$}}
\newcommand{\lapproxeq}{\lower .7ex\hbox{$\;\stackrel{\textstyle <}{\sim}\;$}}

\begin{document}
\thispagestyle{empty}
\begin{flushright}
May 1993 \\
OUTP 93-10-P
\end{flushright}

\bigskip
\begin{center}
{\large {\bf HALF-STRING APPROACH TO CLOSED
           STRING FIELD THEORY}}\\

\bigskip
\medskip

{\large  F. Anton, A. Abdurrahman\footnote{Also at Mathematical
Institute, 24-29, St. Giles, Oxford, OX1 3LB, U.K.} } \\
\bigskip
{\it Department of Theoretical Physics, University of Oxford} \\
{\it 1 Keble Road, Oxford, OX1 3NP, U.K.} \\~\\
{\it and} \\~\\
{\large J. Bordes}\footnote{Also at IFIC, Centro Mixto Universitat de
Valencia-CSIC} \\~\\

{\it Departament de F\'{\i}sica Te\'orica, Universitat de Valencia} \\
{\it Dr. Moliner 50, E-46100, Burjassot, Spain.}

\end{center}

\centerline{ABSTRACT}

\begin{quote}

In this letter we present an operator formalism for
Closed String Field Theory based on closed half-strings.
Our results indicate that the restricted polyhedra of
the classical non-polynomial string field theory,
can be represented as traces of infinite matrices, with
operator insertions that reparametrise the half-strings.
\end{quote}
\begin{center}
\end{center}
\newpage
\vspace{10mm}
The problem of formulating a covariant closed string field theory
proved to be surprisingly difficult.
A straight forward extension of Witten's open string field theory
\cite{WT1} to closed strings does not satisfy gauge
invariance\cite{Giddins} and it fails to reproduce the correct
dual scattering amplitudes \cite{kaku,constraints}.

It is now well established that a consistent closed string field
theory, is necessarily nonpolynomial\cite{zwiebach1}.
At the classical level, the interaction terms
are the so-called {\it restricted}
polyhedra\cite{Saadi,kugo1,kugo2}, where the contact
interactions are given by the patterns of string overlaps on
the surface of a sphere having always three edges at each
vertex. For the N-string scattering one has an N-faced
polyhedron for which the N closed strings
correspond to the faces, and are glued together across the
edges. The lengths of the edges play the role of the modular
parameters. They are restricted by ({\it i}), the
sum of the lengths of the edges of any face equals $2\pi$
(corresponding to the fact that all strings are taken to
have the same $\sigma$-length equal to $2\pi$), and ({\it ii}),
any closed path surrounding two or more faces on a polyhedron
has length larger than or equal to $2\pi$.

On the other hand,
based on the strong analogies between Yang-Mills theory and Witten's
open string field theory, it was first suggested in\cite{CHAN1,CHAN2}
and proved rigorously in~\cite{B1,M1,M2},
that physical open strings can be viewed
as infinite matrices. If one breaks the open string into two pieces,
the string field $\Psi$,
can be treated as a functional of the  two "half-strings",
which play the role of the row and column indices, and a
function of the mid-point. In particular the open string 3-vertex
is represented as a trace\cite{B1,M1,M2}. This trace can be
generalized to represent
any N-string ($N \geq 3$) tree level scattering amplitude\cite{CHAN3}:
\be
A_N = \int_{- \infty}^{ \infty} {d\lambda _1 \cdot \cdot \cdot
d\lambda _N
\over SL(2,\Re )} Tr (\exp (\lambda _1M) A_1 \cdot \cdot \cdot
\exp (\lambda _NM) A_N).
\label{amplitude-trace}
\ee
The operator $M$ is the generator of infinitesimal shifts of the
mid-point of the string, so in fact we are shifting this point
to every possible position.

In an analogous construction for closed strings one
brakes the closed string into two
pieces, therefore singling out two points, and treats each of
the remaining halves as labeling matrix indices. Shifting these two
points around corresponds to varying the lengths of the basic overlaps
of the strings. In other words, one is varying the lengths of the
edges of the polyhedra. Using a functional approach
it has been proved  in\cite{CHAN4}, that the analogous
of eq. (\ref{amplitude-trace}) for closed strings gives the correct
dual amplitudes. Alternatively, one could restrict the region of
integration in such a way that the restricted polyhedra are obtained.
For the 3-vertex there are no modular parameters, so
the Witten vertex should work. For higher vertices
the proper region of integration has to be found in such a way that
the moduli spaces of the field theory are covered correctly.

In this letter we present an oscillator construction for orbital
degrees of freedom of the closed strings based on half-string
coordinates (the ghost degrees of freedom will be treated elswere).
We will find that the
correspondence between the half-string and the full-string
descriptions is non-singular. As examples
we calculate explicitly the $1$ and the
$2$-vertices, and we recover the correct vertex for the scattering of
three tachyons.
We finish this letter by showing how a vertex with no operator insertions
obtained by sewing two vertices, satisfies
the ``gluing and resmoothing theorem'' of \cite{leclair1}
in a trivial way. Our approach will follow closely the approach of
references \cite{B1,B2,M1} and \cite{M2}.

\underline{\hbox{Half-string coordinates}}. The boundary
condition satisfied by the closed string is
(the space-time index is not written, and the
length of the strings is taken to be $\pi$ as opposed to $2\pi$):
\be
X(\sigma ,\tau ) = X(\sigma +\pi ,\tau ).
\label{1}
\ee
The general solution to the string equations of motion compatible
with (\ref{1}) at $\tau=0$, can be written as
\be
X(\sigma ,0) = x_0 + {1\over \sqrt 2} \sum_{n \geq 1} [x_ncos2n\sigma
+y_nsin2n\sigma ],
\label{1.3}
\ee
where we have introduced the oscillators
\begin{eqnarray}
x_n & = & {i \over \sqrt{2}n}(\alpha_n -\alpha_{-n}+\widetilde{\alpha}_n
-\widetilde{\alpha}_{-n}) = x_n^{ \dagger },\\
y_n & = & {1 \over \sqrt{2}n}(-\alpha_n -\alpha_{-n}+\widetilde{\alpha}_n
+\widetilde{\alpha}_{-n}) = y_n^{ \dagger }.
\label{1.2}
\end{eqnarray}
The oscillator modes $\alpha$ and $\widetilde {\alpha}$, satisfy the
well known commutation relations
\begin{eqnarray}
[\alpha _n, \alpha _m] = [\widetilde { \alpha }_n,\widetilde { \alpha
}_m] = m\delta _{n+m}; & [\alpha _n,\widetilde{ \alpha }_m] = 0.
\label{2}
\end{eqnarray}
The Fock space of the theory is constructed by acting
on the vacuum $|0>$,  with the creation operators
$\alpha _{-n}$ and $\widetilde{ \alpha }_{-n}$, (with $n \geq 1$).

There are several possible boundary conditions
for the half string coordinates. The choice we make, will determine
the commutation relations of the half-string oscillator
modes after quantization and therefore the half-string Fock spaces.
The correct choice of boundary conditions  is the one
that isolates completely the motion of the two points
$X(\sigma = {\pi \over 2})$ and $X(\sigma = 0)$, where
we break the string. This is so that there is
a one-to-one correspondence
between the two descriptions (which in general is given by
an non-singular infinite dimensional matrix).
\begin{figure}
\newcounter{cms}
\setlength{\unitlength}{1mm}
\begin{picture}(150,70)
\put(65,20){\oval(46,30)}
\put(52,20){\vector(0,0){15}}
\put(70,20){\vector(0,-1){15}}
\put(42,20){\line(1,0){46}}
\put(60,30){\makebox(0,0){$\chi^{(1)}(\sigma )$}}
\put(79,10){\makebox(0,0){$\chi^{(2)}(\sigma )$}}
\put(35,23){\makebox(0,0){\small{$X(0)$}}}
\put(35,17){\makebox(0,0){\small{$X(\pi)$}}}
\put(94,20){\makebox(0,0){\small{$X(\frac {\pi}{2})$}}}
\put(65,20){\vector(1,1){22}}
\put(95,48){\makebox(0,0){\small{$\frac {2\sigma}{\pi}(X(\frac {\pi}{2})
-X(0))+X(0)$}}}
\label{halfcoord}
\end{picture}
\caption{\it Definition of the Half-String Coordinates}
\end{figure}
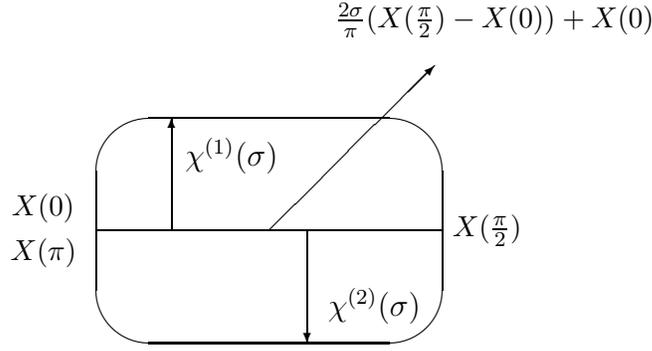
The simplest  choice is the
one depicted in figure (1), where we joined the
string points at $\sigma = {\pi \over 2}$, and $\sigma = 0$ with a
straight line, and the half string coordinates $\chi ^r(\sigma )$, $r=1, 2$,
are defined as the distances between this axis and the string.
One has:
\be
\chi ^r(\sigma ,\tau ) = \left\{ \begin{array}{ll}
X(\sigma ,\tau ) - {2 \over \pi }({\pi \over 2} - \sigma )X(0,\tau ) -
{2\sigma \over \pi }X({\pi \over 2},\tau ) & \mbox{if r=1,} \\
X(\pi - \sigma ,\tau ) - {2\over \pi }({\pi \over 2} - \sigma )X(0,\tau ) -
{2\sigma \over \pi }X({\pi \over 2},\tau ) & \mbox{if r=2.}
         \end{array}
\right.
\label{1.1}
\ee
where $\sigma \in [0,{\pi \over 2}]$.
The boundary conditions are $
\chi ^r (0) = 0 = \chi ^r ({\pi \over 2})$,
and they imply
a Fourier decomposition (at $\tau=0$) in terms of even
sine modes only:
\be
\chi ^r(\sigma ) = \sum_{n \geq 1} \chi ^r _nsin2n\sigma ,
\label{1.4}
\ee
with inverse:
\be
\chi ^r _n = {4\over \pi} \int_{0}^{\pi \over 2} d\sigma \chi^r
(\sigma )sin2n\sigma .
\label{1.5}
\ee
Let us define the following
quantities in the space of ``even-odd oscillators'':
\begin{eqnarray}
{\bf A} ^r _n = \left( \begin{array}{c}
                         A^r_{2n} \\  A ^r_{2n-1}  \end{array}
\right);  & {\bf B}_n =  \left( \begin{array}{c}
                         B_{2n} \\  B_{2n-1}  \end{array}
\right),
\label{1.8}
\end{eqnarray}
Where $A$ ($B$) is any half-string (full-string) vector.
We also define the matrix:
\be
{\bf B}_{n,m} = \left( \begin{array}{cc}
                           \left( {2m-1}\over {2n} \right)B_{2n,2m-1}
& 0 \\ 0 &  \left( {2m}\over {2n-1} \right)B_{2n-1,2m} \end{array}
\right),
\label{1.8b}
\ee
(bold-faced quantities will be used to represent
vectors or matrices in this "even-odd" space).
$B_{m,n}$ is given by:
\be
B_{m,n} = {1\over \pi } \left({2n} \over {n^2 - m^2} \right).
\label{1.7}
\ee
Then one can express $\chi^r_n$ as:
\load{\normalsize}{\bf}
\begin{eqnarray}
\mbox{
\boldmath $\chi $} ^r _n = -\sqrt 2 \sum_{m \geq 1} {\bf B}_{n,m}
                           {\bf u}   {\bf x}_m
                       + {(-)^{r+1}\over \sqrt 2}  {\bf y}_n;  &
{\bf u} = \left( \begin{array}{cc}
            0 & 1 \\ 1 & 0 \end{array} \right).
\label{1.9}
\end{eqnarray}
The matrix ${\bf B}_{n,m}$ is non-singular.
Indeed it can be checked that:
\be
({\bf B}^{-1})_{m,n} = -4 \left( \begin{array}{cc}
                           \left( {2n}\over {2m-1} \right)^2B_{2m-1,2n}
& 0 \\ 0 &  B_{2m,2n-1} \end{array}
\right).
\label{1.10}
\ee
Therefore there is an invertible
relation between the full-string oscillators $x_n$
and $y_n$, and the half-string modes $\chi ^r _n$,
the inverse of equation (\ref{1.9}) is:
\begin{eqnarray}
{\bf x}_m = {-{\bf u} \over 2\sqrt 2} \sum_{n,r} ({\bf B}^{-1})_{m,n}
\mbox{\boldmath $\chi $} ^r _n; &
{\bf y}_n = {1\over \sqrt 2} \sum_{r=1,2} (-)^{r+1}
\mbox{\boldmath $\chi $} ^r _n.
\label{1.11}
\end{eqnarray}
To complete the picture, we still need an explicit relation between
the center of mass coordinate $x_0$, and the two points of the string that
we singled out. Calling them $X_I = X(\sigma = {\pi \over 2},\tau =0 )$
and $X_{II} = X(\sigma = 0,\tau =0 )$, it is straight forward to
see that:
\be
x_0 = {1\over 2}(X_I + X_{II}) + {1\over \pi} \sum_{n,r}
                                             { \chi ^r_{2n-1} \over
                                                            {2n-1}}.
\label{1.12}
\ee
Hence there exists a one-to-one correspondence between the half-string
and the full string pictures. To establish this correspondence
at the level of Fock spaces, we proceed to quantize the theory.

The quantization programme can be carried out in the usual way
by  interpreting
the oscillator modes $\chi^r_n$,  $X_I$ and  $X_{II}$ as q-operators and
define their conjugate momenta as $P_I=-i{\partial \over \partial X_I}$,
$P_{II}=-i{\partial \over \partial X_{II}}$ and
$P^r_n=-i{\partial \over \partial \chi^r_n}$. They satisfy
$$
[\chi^r_n,P^s_n] = i \delta^{r,s}\delta_{n,m},
\, \, \, \, \, [X_I,P_I] = i,
\, \, \, \, \, [X_{II},P_{II}]=i.
$$
The corresponding momenta for the full string are given by
$P_n=-i{\partial \over \partial x_n}$,
$\widetilde{ P}_n=-i{\partial \over \partial y_n}$ and
$P=-i{\partial \over \partial x_0}$.

The transformation rules between the half and the full-string
momenta can be easily worked out using the chain rule to give:
\be
{\bf P} ^r _m = -{{\bf u}\over {2\sqrt 2}} \sum_{n \geq 1}
 ({\bf B}^{-1})_{n,m} {\bf P}_n
                       + {(-)^{r+1}\over \sqrt 2} {\bf \widetilde{ P}}_m
                       + {{\bf e}_2 \over \pi} {P\over (2m-1)}
\ee
and
\be
{1\over 2}P=P_I +P_{II}.
\label{2.4}
\ee
Here ${\bf e}_2$ is the unit vector $(0,1)^T$. These equations can be
inverted to solve for $P_m$,
\begin{eqnarray}
{\bf u}{\bf P} _m = -\sqrt 2 \sum_{n,r} {\bf B}_{n,m} {\bf P} ^r_n +
                {{\bf e}_2 \over \sqrt 2} P & \mbox{and} &
{\bf \widetilde{ P}}_m = \sum_r
{(-)^{r+1}\over \sqrt 2} {\bf P} ^r_m.
\label{momenta}
\end{eqnarray}
$P$ is recovered by multiplying the first of eqs. (\ref{momenta})
by ${1\over 2k-1}$, and
summing over $r=1,2$ and odd $n \geq 1$:
\be
P = {4\over \pi} \sum _{r,k} {1\over 2k-1} P^r_{2k-1}.
\label {pzero}
\ee
Relation (\ref{2.4}) may seem somewhat obscure at this point. It
can be better understood at the classical level in the following
way: from equation (\ref{1.1}) we can get an expression for the
string Lagrangian, from which $P_I + P_{II} =
\int _0^{\pi \over 2} \sum _r P^r$, follows as a primary constraint on
the system. The total momentum of the string  is
$P_{\it {total}} = P_I + P_{II} + \int _0^{\pi \over 2} \sum _r P^r =
2(P_I + P_{II})$, in agreement with equation  (\ref {2.4}).
This identification means that
$2(P_I+P_{II})$ corresponds to the translational mode of the string. This
will become clear since integration over the combination
${1\over 2}(X_I+X_{II})$ gives rise to the conservation of
momentum in the vertices.

Let us pause for a moment too see the
meaning of this. Traditionally, people have been bias against using
half-strings to describe closed strings
because this would produce a bilocal
theory. However from eqs. (\ref{1.12}) and (\ref{2.4}) we see that in
fact the string fields only depend on the two preferred
points through the combination $X_I+X_{II}$.

The full-string Fock space is build up by acting on the vacuum $|0>$ with
the creation operators:
\begin{eqnarray}
a^{\dagger}_m =
{1\over \sqrt {2m}}(\alpha _{-m} + \widetilde {\alpha }_{-m}); &
\widetilde {a}^{\dagger}_m = {i\over  \sqrt {2m}}(\alpha _{-m} -
\widetilde {\alpha }_{-m}),
\end{eqnarray}
corresponding to the modes $x_n$ and $y_n$ respectively.
The creation  and annihilation operators for the
half-string defined as:
\begin{eqnarray}
{\beta ^r_n}^{\dagger } =
i {\sqrt n \over 2}\chi ^r_n + {1\over \sqrt n} P^r_n; &
\beta ^r_n =  -i {\sqrt n \over 2}\chi ^r_n + {1\over \sqrt n} P^r_n,
\label {creophalf}
\end{eqnarray}
are related to the full string oscillator modes by
\begin{eqnarray}
{\bf u}{\bf a}_m = \sum _{n \geq 1} ({\bf A}^{(+)}_{mn}
                            {\bf b}_n^{(+){\dagger }}
                 +          {\bf A}^{(-)}_{mn}
                            {\bf b}_n^{(+)})
                 + {{\bf e}_2 P \over 2\sqrt m}; & \mbox{and} &
{\bf \widetilde {a}}_m  = {\bf b}^{(-)}_m,
\label{abetas}
\end{eqnarray}
where  $b^{\pm}_n = {1\over \sqrt 2}
(\beta ^{(1)}_n \pm \beta ^{(2)}_n)$, and
we have defined the
following matrices:
\be
{\bf A}^{\pm }_{m,n} = \left( \begin{array}{cc}
                           -N^{\pm }_{m,n}
& 0 \\ 0 &  -M^{\pm }_{m,n} \end{array}
\right);
\label{Apm}
\ee
\be
M^{\pm }_{mn} = \left({2m\over {2n-1}}\right)^{1\over 2} [B_{2n-1,2m} \pm
                                               B_{2m,2n-1} ];
\label {Mpm}
\ee
and
\be
N^{\pm }_{mn} = \left({2n\over {2m-1}}\right)^{1\over 2}
[\left( {2m-1 \over 2n} \right)B_{2n,2m-1} \pm
                  \left({2n \over 2m-1}\right)B_{2m-1,2n} ].
\label {Npm}
\ee
Notice that the matrix $M^{\pm}$, coupling the odd oscillators already
appeared in \cite {B1,B2} for the open
bosonic string. Next, we go on to develop a relation between the full
string vacuum $|0>$, and the two half-string
vacua $|0>_1$ and $|0>_2$ corresponding to the two half-strings. They
are defined in the usual way:
\be
\beta ^r_m |0>_r =0.
\label {vacua}
\ee
Repeated action of the creation operators $\beta ^{r \dagger } _m$,
on these vacua give the Fock spaces corresponding to each
half-string. We propose the standard ansatz
\be
|0> = \exp(\frac {-1}{2}{\bf b}^{(+) \dagger }_n
                      \mbox{\boldmath $\xi $} _{nm}
                        {\bf b} ^{(+) \dagger}_m) |0>_1|0>_2.
\label {vacuum}
\ee
Notice that only the plus combination appears
in the exponential. This is because $\widetilde {a}_n$ is expressed
only in terms of $b^{(-)}_n$,
and hence it annihilates the vacuum trivially.
The action of $a_n$ fixes $\mbox{\boldmath $\xi $}$ to be:
\be
\mbox{\boldmath $\xi $}
= {\bf A}^{(-)-1} {\bf A}^{(+)} =  \left( \begin{array}{cc}
                           \psi
& 0 \\ 0 &  \varphi \end{array}
\right).
\label {xi}
\ee
The tachyon state is
obtained by inserting the factor $e^{iPx_0}$ in (\ref {vacuum}).
One can verify that this state in the half-string representation is
an eigenstate of the momentum operator
with eigenvalue $P$, when expressed in the half-string language.

The Hilbert space $\cal H$ of the closed full string is spanned by linear
combinations of the state vectors of the form
$\prod {a^{\dagger }_n,\widetilde {a}^{\dagger }_n}
|0>$.  The one-to-one relation between full and half-string modes,
together with equation (\ref {vacuum}) imply
that the Hilbert space of the full string is contained in the completion of
the tensor product $\overline {\cal {H}} =
\overline { {\cal H}_1 \bigotimes {\cal H}_2 \bigotimes {\cal H}_M}$.

\underline{\hbox{Matrix representations and verices}}.
In order to compute the half-string matrix representing a string state
with arbitrary ocupation nubers, it is useful to use the coherent
state:
\be
|\vec {\lambda }, \vec {\widetilde{\lambda }};p) = e^{ipx_0}\exp (
{\lambda \cdot a^{\dagger } + \widetilde{\lambda } \cdot
\widetilde{a}^{\dagger }} )|0>.
\label{cohstate}
\ee
Also one introduces the half-string states:
\be
|n^r_i> = \prod _{i=1}^{\infty } {1\over \sqrt n^r_i!}(\beta
^r_i)^{\dagger n^r_i}|0>_r,
\label{hsstates}
\ee
in terms of which, the matrix representing the functional
for any state is defined by:
\be
[A]^{n^2_i}_{n^1_i} = (-)^{\sum_{i \geq 1}n^2_i}<n^1_i;n^2_i|\lambda ,
\widetilde {\lambda } ; p).
\label{13}
\ee
The factor $(-)^{\sum_{i \geq 1}n^2_i}$ takes into account that
the normalization for $\chi ^2$ is reversed.

After a long calculation we find the following expression for
the matrix (\ref{13}):
\begin{eqnarray}
[A]^{n^2_i}_{n^1_i} & = & C(\lambda ;p)e^{{ip\over
2}(X_I+X_{II})}\prod_{i=1}^{\infty }{1\over \sqrt
{n^1_i!n^2_i!}}\left( {-1\over \sqrt 2}D^-_i \right) ^{n^1_i}\left(
{1\over \sqrt 2}D^+_i \right) ^{n^2_i}  \times \nonumber  \\
                    &   &  \exp {\left( {-1\over 2}
{\bf v}^T\mbox{\boldmath $\xi $} {\bf v}  \right )}
\left.   \right| _{v=0},
\label{HSmatrix}
\end{eqnarray}
where we have defined the quantities:
\begin{eqnarray}
C(\lambda ;p) & = & \exp (-{1\over 2}p^2{\bf k}^T({\bf 1}+
\mbox{\boldmath $\xi $}){\bf k} +
p{{\mbox{\boldmath $\lambda $}}}^T({\bf A}^--{\bf A}^+
\mbox{\boldmath $\xi $}){\bf k} \nonumber \\
   & & +{1\over 2}{\mbox{\boldmath $\lambda $}}^T
({\bf A}^-- {\bf A}^+\mbox{\boldmath $\xi $})
{\bf A}^{(+)T} \mbox{\boldmath $\lambda $}).
\label{Clambdap}
\end{eqnarray}
\begin{eqnarray}
\mbox{\boldmath $\lambda $}_i = \left( \begin{array}{c}
                         \lambda ^{\prime}_i \\  \lambda _i  \end{array}
\right); &  \widetilde {\mbox{\boldmath $\lambda $}}_i =
\left( \begin{array}{c}
                     \widetilde {\lambda }_i      \\
  \widetilde {\lambda } ^{\prime }_i
  \end{array}
\right),
\label{17}
\end{eqnarray}
(lambdas with a prime are those that couple to the odd modes of the
creation operators in equation (\ref{cohstate})).
\begin{eqnarray}
D^{\pm }_i = \rho _i \pm \widetilde {\lambda }_i +{d \over d
v_i}; &
{\bf k}_i ={-1\over {\pi \sqrt 2}}  \left( \begin{array}{c}
                         0 \\  (2i-1)^{-{3\over 2}}  \end{array}
\right).
\label{27}
\end{eqnarray}
$v$ is a dummy variable and $\rho $ is given by
\be
\rho _i^T = [\mbox{\boldmath $\lambda $}^T({\bf A}^-+{\bf A}^+
\mbox{\boldmath $\xi $})]_i -p[{\bf k}^T({\bf 1}+
\mbox{\boldmath $\xi $})]_i.
\label{31}
\ee
The coherent state matrix (\ref{HSmatrix}) can be used to construct the
vertices in the half-string language as:
\be
V_N = {1\over 2}\int d(X_I + X_{II}) \int_{D_N}
{d\lambda _1 \cdot \cdot \cdot d\lambda _N
\over SL(2,\Re )}
 Tr (\exp (\lambda _1M) A_1 \cdot \cdot \cdot
\exp (\lambda _NM) A_N).
\label{closedINTERACTION}
\ee
The operator $M=L_1-L_{-1}+\widetilde {L}_1-\widetilde {L}_{-1}$ is
the element of the $SL(2;\Re)$ subalgebra of the Virasoro algebra that
reparametrises $\sigma$ in such a way as to shrink the segment between
$\sigma=\frac {\pi}{4}$ and $\sigma=\frac {3\pi}{4}$, while expanding
its complement\cite{CHAN4}. Although this operator leaves the points
$\sigma=0,\pi$ and $\frac {\pi}{2}$ invariant, one is free to move
these points to any other point by a rigid rotation since, in order to
build the field theory, these
vertices still have to be multiplied by the projection operator $P$
that ensures that condition (\ref{1}) is satisfied.
$D_{N}$ represents the region in moduli space such the restricted
polyhedra are obtained. We postpone to a later paper its
determination. For $N=1,2$ and $3$ no $M$ insertions are necessary.

As examples we will compute the (integration) 1-vertex, the
$2$-vertex (sewing ket) and the $3$-vertex that couples three
tachyons. Setting
$N=1$ in equation (\ref {closedINTERACTION}), using equation
(\ref {HSmatrix}) gives:
\be
I = \delta (P) C(\lambda ;P) \exp {({-1\over 2} D^+D^-)}
                             \exp {({-1\over 2}{\bf v}^T
                                   \mbox{\boldmath $\xi $} {\bf v})} \left.
                                    \right|_{v=0}.
\ee
Integration over the combination ${1\over 2}(X_I+X_{II})$ gave
rise to the momentum conservation delta. The sum over $n^1_{n_i}$ on the
trace was grouped in the exponential. Using the standard techniques in
Gaussian integrals we can write this as:
\be
I = C(\lambda; P=0)\exp {({1\over 2}[\widetilde
{\mbox{\boldmath $\lambda $}}^2
                         -\mbox{\boldmath $\rho $}
^T({\bf 1}-\mbox{\boldmath $\xi $})^{-1}
                          \mbox{\boldmath $\rho $}])}.
\ee
It is not difficult to check that:
\be
({\bf 1}-\mbox{\boldmath $\xi $})^{-1} = ({\bf A}^- +{\bf A}^+)^T{\bf A}^-,
\ee
so that
\be
I = \exp {(\frac {1}{2}[\widetilde {\mbox{\boldmath $\lambda $}}^2 -
\mbox{\boldmath $\lambda $}^2])}.
\label {identity}
\ee
This vertex precisely corresponds to folding the closed string
along its diameter.

Setting $N=2$ in equation (\ref {closedINTERACTION}) gives
\be
V_2 = \delta (P_1 + P_2) C(\lambda _1;P_1)C(\lambda _2;P_2)
 \exp {({-1\over 2} \vec {D}^+B\vec {D}^-)}
                             \exp {({-1\over 2}{\bf v}^T
                            \mbox{\boldmath $\xi $} {\bf v})} \left.
                                    \right|_{v=0}.
\ee
Where as before integration over the combination ${1\over
2}(X_I+X_{II})$
gives
rise to the momentum conservation delta, and the sum  on the
trace was grouped in the exponential. It is understood that the
above quantities have been imbedded in two dimensional vectors or
matrices. For example $\vec {p} =(p_1,p_2)$
are the momenta of the two external states, etc. The matrix $B$ is defined
as
\be
B = \left( \begin{array}{cc}
                       0 & 1 \\
                       1 & 0 \end{array} \right).
\label {B}
\ee
In the same way as before, we can write $V_2$  (ignoring the momentum
conservation delta) as:
\be
V_2 = C(\lambda _1; P_1)C(\lambda _2; P_2)
\exp {({1\over 2}[\widetilde {\mbox{\boldmath $\lambda $}}^T
                 B\widetilde {\mbox{\boldmath $\lambda $}}
                         - \mbox{\boldmath $\rho $}^T
                        (B-\mbox{\boldmath $\xi $})^{-1}
                           \mbox{\boldmath $\rho $}])}.
\ee
It is not difficult to check that:
\be
-\frac {1}{2}\mbox{\boldmath $\rho $} ^T
(B-\mbox{\boldmath $\xi $ })^{-1}\mbox{\boldmath $\rho $}
 = -{1\over 2}\mbox{\boldmath $\rho $}_1{\bf A}^{+T}
        {\bf A}^-\mbox{\boldmath $\rho $}_1
    -{1\over 2}\mbox{\boldmath $\rho $}_2{\bf A}^{+T}
          {\bf A}^-\mbox{\boldmath $\rho $}_2
    -\mbox{\boldmath $\rho $}_1 {\bf A}^{-T}{\bf A}^-
\mbox{\boldmath $\rho $}_2
\ee
so we arrive at
\be
V_2  = \exp {(\widetilde {\mbox{\boldmath $\lambda _1 $}}^T \cdot
              \widetilde {\mbox{\boldmath $\lambda _2 $}}
 - \mbox{\boldmath $\lambda _1 $}^T \cdot
\mbox{\boldmath $\lambda _2 $})},
\label {V2}
\ee
which again can be seen to correspond to the standard result.

We now proceed to the computation of the 3-vertex.
{}From (\ref {closedINTERACTION}):
\begin{eqnarray}
V_3 & =  \delta (\sum_{i=1}^{3}p_i)e^{(a\sum_{r=1}^3p^{(r)^2})} \nonumber \\
        & \exp [\widetilde{ \vec{ \lambda }}
^TM_1\widetilde{ \vec{ \lambda }}+\vec{ \lambda}^TM'_1\vec{ \lambda }
        + \widetilde{ \vec{ \lambda }}^TM_2\vec{ p}+
\vec{ \lambda }^TM'_2\vec{ p}+
\widetilde {\vec{ \lambda }}^TM\vec{ \lambda }].
\label{34}
\end{eqnarray}
As before we have imbedded the above quantities in the
three dimensional space spanned by the three strings.
The quantity $a$ and the matrices $M_i$, $M'_i$ and $M$,
are the same quantities that appeared in \cite{B1}, with
$\phi \rightarrow \mbox{\boldmath $\xi $} $
and $k_n \rightarrow {\bf k}_n$.

Setting $\lambda = \widetilde
{\lambda } = 0$ in the above expression, one recovers the
interaction of three tachyons\footnote{Recall that
for closed strings $P^2=\frac {1}{2}P^2_{left}+
\frac {1}{2}P^2_{right}=\frac {1}{2}P^2_{open}.$},
namely
\be
V_3^{\it Tachyons} = \delta (\sum _{i=1,2,3} P_0^{i})
         e^{\frac{-1}{8}ln(\frac{3^3}{2^4})\sum_{i=1,2,3}P^{(i)2}_0}.
\label {3tach}
\ee
In arriving at this expression we have used that $a=-{1\over 4}N^{rs}_{00}$,
which can be easily established with the results of references\cite {B1,B2}.

These three examples coincide with the standard expressions. However
we still have to compare our higher order
vertices with other vertices appearing
in the literature. In \cite{leclair1}, LeClair, Peskin and Preitschopf
(LPP) define the 3-vertex through the relation:
\be
<V_{123}||A>_1|B>_2|C>_3=<T^2h[\Theta _A]Th[\Theta _B]h[\Theta _C]>,
\label{LeClair's}
\ee
here $\Theta _A$ is the normal ordered operator that
creates the state $|A>$:
\be
|A>=\Theta _A|0>,
\ee
and $T$ is an $Sl(2;C)$ transformation such that $T^3=1$.
Higher order vertices are defined analogously.

There are precisely
these type of vertices the ones used as a starting point
by Kugo and Suehiro, to
construct the restricted polyhedra, and to show that the resulting
theory is gauge invariant\cite{kugo2}.

Lets start by comparing the two-point vertices in both formalisms.
By definition the state $|V_{12}>$ (or simply
$|V^{(2)}>$ in our language) imposes the condition
$X_1(\sigma)=X_2(-\sigma )$, which is equivalent to imposing
that $\alpha ^{(1)}_n =\alpha ^{(2)}_{-n}$ at the level of operators.
If we write the string coordinate in terms of the complex coordinate
$z=\exp (\tau +i\sigma)$ (take $\tau =0$), then this condition is
equivalent to imposing $X_1(z)=X_2(\frac{1}{z})$. This can be used
to
define the Belavin,
Polyakov and Zamolodchikov (BPZ) conjugate to $|A>$ as
\be
<A| = <0| I[\Theta _A(0)],
\label{BPZconj}
\ee
where $I(z)=\frac{1}{z}$ is a conformal map. i.e.:
\be
I[A(z,z^*)]=A^{\prime}(\frac{1}{z},\frac{1}{z^*}).
\ee
The (BPZ) inner product is defined as:
\be
<A|B>=<0|I[\Theta _A(0)]\Theta _B(0)|0>=<V_{AB}||A>|B>.
\label{innerproduct}
\ee
It is easy to see that the $2$-vertex constructed with LPP's prescription
using the $I$ defined above, corresponds exactly to our $V_2$.
\par
A very important property any vertex should satisfy is that the new
vertex produced by the contraction of two other vertices by the
conformal field theory inner product  BPZ, is precisely the one
which results from first sewing the corresponding
two Riemann surfaces via the map
$I$, and then constructing the vertex on that
surface. In other words:
\be
<V^{(4)}_{ABEF}| = <V^{(3)}_{ABC}|<V^{(3)}_{DEF}||V^{(2)}_{CD}>,
\label{sewing}
\ee
with
\be
<V_{1234}||A>_1|B>_2|C>_3|D>_4
=<T^2h[\Theta _A]Th[\Theta _B]IT^2[\Theta _C]IT[\Theta _D]>.
\ee
This is the {\it Generalized  Gluing and
Resmoothing Theorem} (GGRT) for $N=3$ \cite{leclair1}.

It is straight forward to show that our vertices satisfy this
theorem. The proof steams out from the fact that they are written as
traces, and that the transformation from half-strings to
full-strings is complete.

Denote our string field matrices as $Anm=<nm|\Lambda)$ where the
indices $n$ and $m$ refer to the left and right parts of the string,
and  $\Lambda $ is a short for $(\lambda , \widetilde
{\lambda })$. Completeness of the (orthogonal) transformations means
that Parseval's identity, $I=\sum_{nm} |n;m><n;m| $, works both ways,
and $I=\int D\Lambda |\Lambda)(\Lambda|$.
The left hand side of equation (\ref{sewing}) can be written in our
language as (summing over repeated indices):
\begin{eqnarray}
V^{(3)}_{125}V^{(2)\dagger}_{56}V^{(3)}_{634}  =
\int D\Lambda ^5 D\Lambda ^6 <nm|\Lambda ^1)<mk|\Lambda ^2)<kn|\Lambda
^5) \nonumber \\
(\Lambda ^5|pq>(\Lambda ^6|qp>
<rs|\Lambda ^6)<sv|\Lambda ^3)<vr|\Lambda ^4).
\end{eqnarray}
Using (twice) that $\int D\Lambda <kn|\Lambda)(\Lambda|pq> =
\delta_{kp} \delta_{nq}$ (orthogonality), we can write the above
equation as
\be
V^{(4)}_{1234}= <nm|\Lambda ^1)<mk|\Lambda ^2)<kq|\Lambda^3)
<qn|\Lambda^4),
\ee
which is just the l.h.s. of (\ref{sewing}).

This can be generalized trivially to higher point vertices. Notice
that this result is valid for both open and closed strings.

\underline{\hbox{ Conclusions}}.
We have showed that closed string fields
can be represented by matrices, in a very similar way as the
open string ones. We only dealt with the orbital degrees of freedom,
but the ghosts can also be treated with half-strings in a very
similar way as in references \cite{M1,M2}.
We defined the closed half-string coordinates,
and  from them we constructed the corresponding half-string matrices.
These are
written in a very similar way as the matrices corresponding to open
string functionals. We constructed the vertices $V_N$ for $N=1,2$ and
$3$ explicitly, and we showed that --as for open strings-- they are
determined completely by the matrix  ${\bf B}_{nm}$, which
relates the full and the half string modes. This will still hold for
higher order interactions. We obtained the correct vertex
that couples three tachyons together
and showed that our vertices satisfy the ``general gluing and
resmoothing theorem'' of \cite{leclair1} in an almost trivial
way.

\enddocument